# From Individual Songs to Music Corpora: Evaluating Explainable Harmonic Decision-Making on HookTheory

**Josef Pavlíček[1] and Martin Pelikán[2]**

[1] Faculty of Economics and Management, Czech University of Life Sciences in Prague, Prague 6, 165 00 Praha-Suchdol, Czech Republic, pavlicek@pef.czu.cz

[2] Faculty of Economics and Management, Czech University of Life Sciences in Prague, Prague 6, 165 00 Praha-Suchdol, Czech Republic, pelikan@pef.czu.cz

**ABSTRACT**

Automatic melody harmonization is often evaluated through individual compositions or data-driven corpus models. This study examines whether an explainable harmonic decision model developed for individual songs can be applied to independently collected human annotations without changing its rules. We evaluate a frozen Explainable Melody Harmonizer (EMH) on selected HookTheory excerpts containing aligned melody and chord annotations. EMH generates chord sequences from melody alone using explicit criteria for melodic compatibility, harmonic function, persistence, cadence, meter-aware segmentation, and Viterbi-like optimization. Human chords are withheld during generation and used only for post-hoc comparison.

Nineteen excerpts yielded 439 evaluable harmonic segments. Exact chord agreement reached 20.50%, chord-root agreement 31.66%, and harmonic-function agreement 47.62%, with functions assessed relative to the key inferred by EMH. Decision-level analysis identified 203 segments (46.24%) where the human chord lay outside the model's candidate vocabulary and 146 (33.26%) where it was available but not selected. These findings distinguish representational limitations from differences in chord selection without treating the human reference as the only musically valid solution. The study demonstrates a reproducible, interpretable approach to corpus-based evaluation of harmonic decision-making.

## INTRODUCTION

A single melody admits multiple plausible harmonizations. Accordingly, agreement with an annotated chord progression is not equivalent to musical quality, nor does melodic compatibility establish that an algorithm has reconstructed the original composer’s decisions. Computational studies of harmony document regularities in chord organization and listener expectations (Moss et al., 2019; Miles et al., 2017), while theories of tonal cognition emphasize contextual and hierarchical interpretation (Lerdahl and Jackendoff, 1983; Krumhansl and Carol, 2001). These observations motivate evaluation at the level of interpretable decisions rather than treating every difference from a reference as an undifferentiated error.

Our preceding case-based study compared six narrative-associated songs with computational harmonizations and showed how harmonic persistence, functional organization and non-diatonic events could be preserved or transformed despite high internal melody–chord compatibility (Pavlíček et al., 2026, prior AHFE study). The present work moves from individual compositions to independently collected corpus excerpts. We retain the same frozen Explainable Melody Harmonizer (EMH v0.12) and evaluate it on a selected HookTheory subset without adjusting its decision rules to the reference harmonies.

The research questions are:

- **RQ1**—To what extent does a frozen explainable harmonizer reproduce human chord decisions on independently collected excerpts?
- **RQ2**—How much observed disagreement reflects unavailable reference chords, and how much persists when a reference chord is selectable?
- **RQ3**—What interpretive information does a decision-level diagnostic provide beyond overall chord accuracy?

Our contribution is a reproducible corpus-evaluation procedure with a strict reference-harmony firewall and an explicit decomposition of disagreement.

## BACKGROUND AND RELATED WORK

### Computational harmony and tonal decision-making

Harmony can be described using chord vocabulary, temporal organization, transition patterns and tonal function. Corpus work has characterized statistical properties of tonal chord sequences (Moss et al., 2019), and research on harmonic surprise relates expectation to listener preference (Miles et al., 2017). Foundational accounts of tonal hierarchies and generative musical structure offer complementary cognitive interpretations (Krumhansl et al., 1982; Lerdahl and Jackendoff, 1983). Probabilistic melodic harmonization and automatic chord analysis illustrate that candidate generation and sequence selection are separable computational problems (Temperley, 2006; Sumi et al., 2008).

### Corpus methods and explainability

Large-scale music generation and sequence modeling have advanced through neural architectures, including Transformer-based models, but the availability of a plausible output does not by itself expose why a particular harmonic choice was made (Huang et al., 2018). Explainable AI research distinguishes transparency of a decision procedure from post-hoc explanation and stresses the importance of appropriate interpretation by human users (Adadi and Berrada, 2018; Arrieta et al., 2020). In the musical domain, explicit scoring terms, restricted candidate vocabularies and inspectable paths permit an additional kind of analysis: whether a mismatch arises because a choice was impossible under the model, or because another candidate was preferred.

HookTheory provides aligned symbolic melodic and harmonic annotations, and the publicly distributed SheetSage data release includes a corpus representation and split metadata (Donahue et al., SheetSage data repository). We use these annotations as a comparison reference, not as training data or a uniquely correct musical solution. Earlier research by the authors used BPMN and the Chord Wheel Diagram to represent temporal and tonal-functional aspects of harmonic decisions (Pavlicek et al., 2024; Pavlicek et al., 2025; Pavlíček et al., 2026). The present experiment focuses on algorithmic diagnostics rather than testing the usability of those visualizations.

## EXPLAINABLE MELODY HARMONIZER

### Frozen baseline and candidate vocabulary

EMH (Fig.1) v0.12 processes monophonic melodic events, infers a tonal context from melody alone, constructs meter-aware harmonic decision segments and selects a chord sequence by global dynamic programming. The frozen candidate vocabulary comprises seven diatonic triadic functions (I, ii, iii, IV, V, vi and vii°) and three selected seventh-chord functions (ii7, V7 and viiø7), realized relative to the detected key. It excludes many chromatic, extended and added-tone harmonies, including secondary dominants and ninth, eleventh and thirteenth chords. These exclusions are intentional model properties rather than evaluation failures.

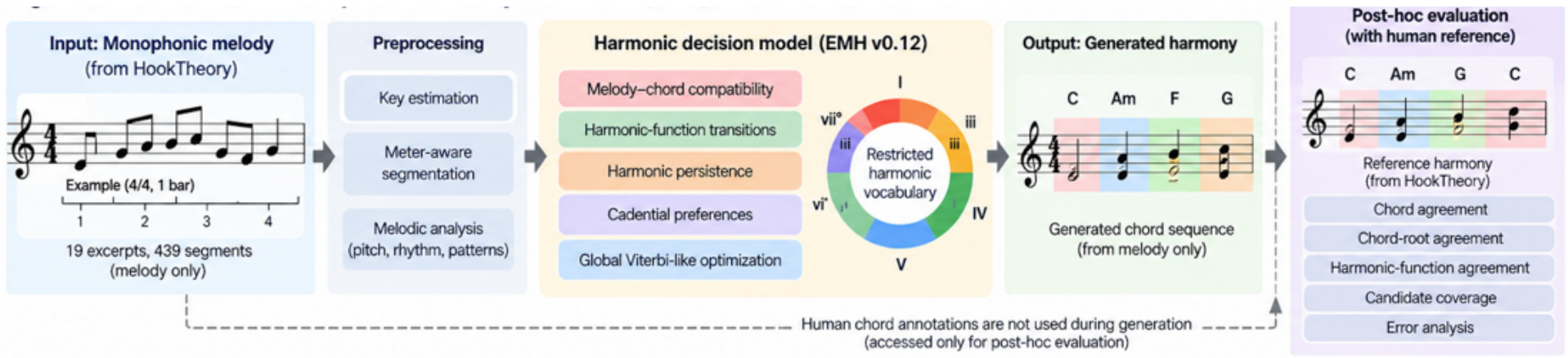

**Figure 1:** Architecture of the Explainable Melody Harmonizer (EMH) and the HookTheory evaluation workflow. The system generates harmony from monophonic melody using explicit harmonic criteria and global optimization. Human chord annotations are excluded from generation and used exclusively for post-hoc evaluation. Illustration prepared with generative AI assistance.

### Transparent scoring and path optimization

The system scores melody–chord compatibility, functional transitions, harmonic persistence and cadential preferences using fixed weights of 0.50, 0.25, 0.10 and 0.15, respectively. It then identifies the best complete path with a Viterbi-like dynamic-programming procedure. Meter-aware segmentation defines the available decision positions. This formulation is globally optimized, so a locally favorable chord need not appear in the final sequence. The algorithm and its parameters were not modified for the corpus experiment.

An inspectable decision record contains the selected chord, its function and scoring information, together with alternatives. These records support analysis of available versus unavailable reference choices; however, the presence of a different selected chord does not alone establish which particular scoring term caused the difference.

## MATERIALS AND METHODS

### Dataset selection and experimental firewall

The corpus was obtained from the publicly distributed HookTheory JSON archive. A deterministic selection manifest identified 20 candidate excerpts from the dataset TEST partition, subject to melodic and metrical eligibility criteria. Nineteen excerpts were processed; one was excluded because compound-meter time-base mapping had not been verified in the adapter. The present evaluation is a proof of concept rather than a representative population estimate. Selection, exclusion and processing status are recorded in the experiment outputs.

Only melodic events and necessary structural timing information are passed to the frozen generator. Human chord annotations are accessed separately, after harmonization, by the evaluator. This firewall prevents direct leakage of reference chord labels into chord generation. The detected key is obtained from the melody rather than from the human chord sequence. All comparisons use the same unmodified candidate set and weights.

### Alignment and evaluation measures

Generated harmony is evaluated on the fixed EMH decision segments. A reference chord is assigned to each segment by maximum temporal overlap with the human annotations. This convention enables a common comparison grid but may simplify reference segments containing multiple chords. Forty evaluated segments contain mixed reference harmony; four have incomplete reference coverage. These conditions are retained as diagnostic limitations rather than silently discarded.

We report chord-root agreement; chord agreement using the evaluator's normalized chord representation; functional agreement under the tonal context detected by EMH; candidate coverage, which tests whether the reference chord belongs to the model's available vocabulary; root candidate coverage; agreement conditional on candidate coverage; and root agreement on unmixed segments. Functional agreement is not based on independently annotated human functional labels. All aggregate rates are micro-averages over eligible evaluated segments.

### Reproducibility

The experiment was executed with the tagged repository release emh-v0.12.1-hooktheory. The original EMH implementation remains unchanged; separate HookTheory loader, adapter, evaluator and audit scripts produce machine-readable summaries. The diagnostic audit partitions each evaluated segment into exact match, reference unavailable or reference available but not selected. Code and manifest are available at https://github.com/JosefPavlicek/quantum-inspired-music-research (tag emh-v0.12.1-hooktheory). The external dataset is identified by its original provenance and license.

## RESULTS

The 19 processed excerpts yielded 439 evaluated harmonic segments. One of the 20 selected cases was excluded for unverified compound-meter mapping. Table 1 summarizes the agreement and vocabulary-coverage measures. Because denominators differ, especially for functional agreement, the measures should not be treated as interchangeable accuracies (Fig.2).

**Table 1.** Corpus-level agreement and candidate coverage for frozen EMH v0.12

| Measure | Matches / evaluated | Rate |
|---|---|---|
| Chord-root agreement | 139 / 439 | 31.66% |
| Exact chord agreement | 90 / 439 | 20.50% |
| Harmonic-function agreement | 200 / 420 | 47.62% |
| Candidate chord coverage | 236 / 439 | 53.76% |
| Candidate root coverage | 420 / 439 | 95.67% |
| Chord agreement \| candidate covered | 90 / 236 | 38.14% |
| Root agreement, unmixed segments | 125 / 399 | 31.33% |

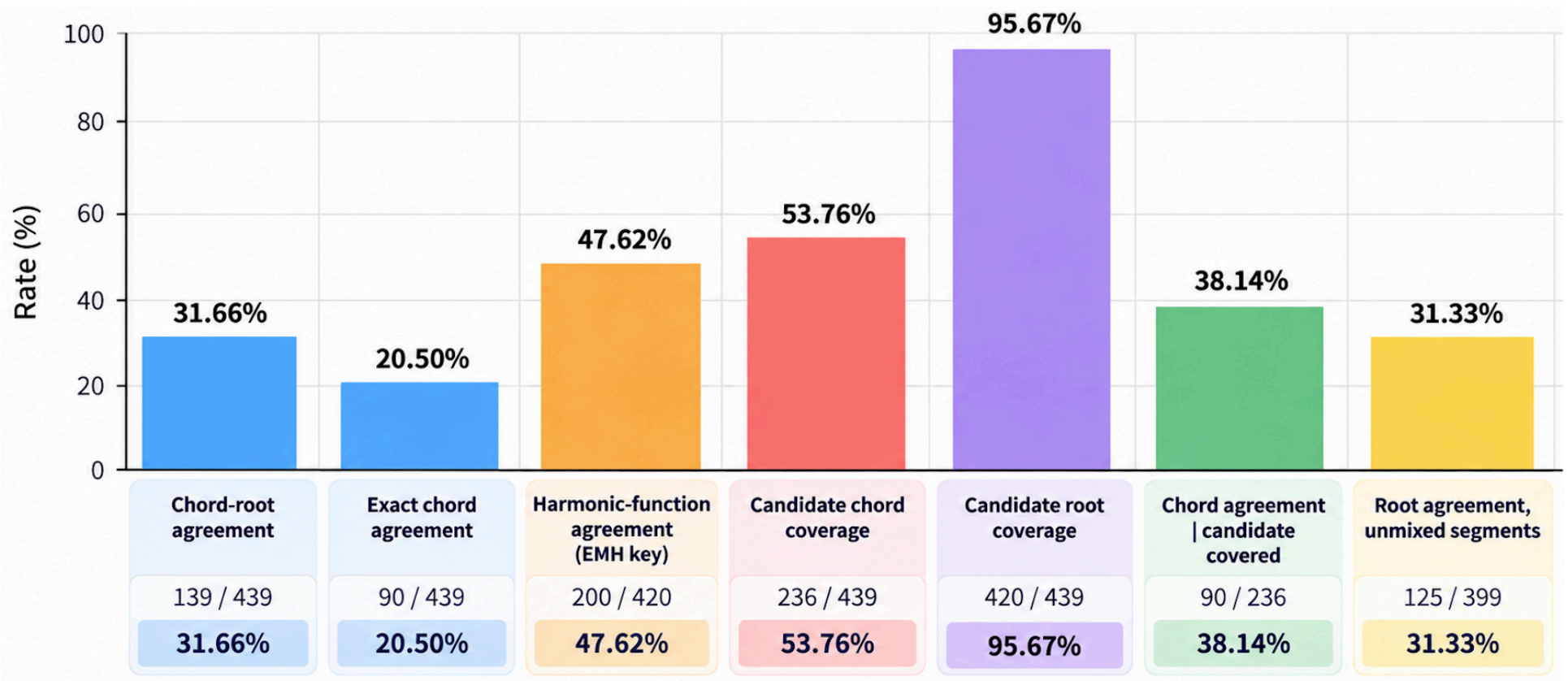

**Figure 2:** Evaluation metrics for generated and human harmony. Bars indicate agreement or candidate coverage rates, with corresponding numerators and denominators. Harmonic-function agreement is evaluated relative to the key inferred by EMH.

The principal decision-level audit yields three mutually exclusive categories (Table 2). Ninety segments exactly match the human chord (20.50%). In 203 segments (46.24%), the reference chord is absent from the candidate vocabulary and thus cannot be reproduced exactly. In another 146 segments (33.26%), the reference is available but a different chord is selected. These categories sum to all 439 evaluated segments (Fig.3).

**Table 2.** Mutually exclusive decision-level outcomes.

| Outcome | Segments | Share |
|---|---|---|
| Exact chord match | 90 | 20.50% |
| Human reference absent from candidate set | 203 | 46.24% |
| Reference available but not selected | 146 | 33.26% |
| **Total** | **439** | **100.00%** |

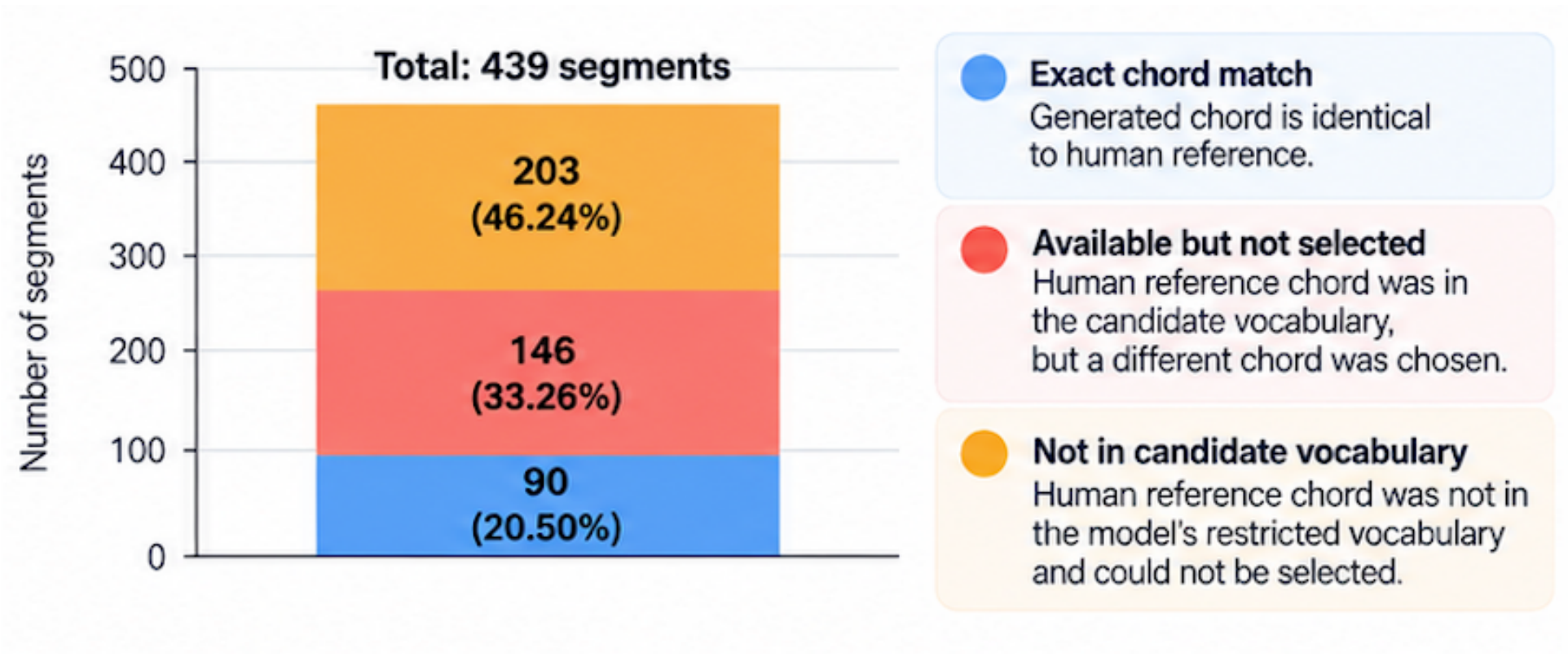

**Figure 3:** Decomposition of harmonic decisions (n=439 segments)

The difference between root coverage (95.67%) and full chord coverage (53.76%) is consistent with substantial restrictions on chord quality and extension rather than root availability alone. Even among segments where the reference chord is representable, exact selection occurs in only 38.14%. Functional agreement (47.62%) exceeds exact chord agreement, suggesting that function-level correspondence can remain when specific chords differ; this remains conditional on the EMH-derived tonal context.

## DISCUSSION

### Separating representation from selection

**RQ1** is answered descriptively by the agreement measures. Exact chord agreement is modest, but the model was deliberately frozen and the human reference is one possible harmonization, not a unique optimum. The more informative result for **RQ2** is the decomposition into 203 structurally unavailable reference chords and 146 available but unselected chords. The former are direct consequences of the restricted candidate representation; the latter identify decisions requiring closer investigation of melodic compatibility, functional context, persistence and global optimization. The current audit does not establish a causal explanation for each individual selection.

### Human-centered interpretation

For **RQ3**, decision-level diagnostics offer a practical advantage over a single similarity score. A musician or researcher can distinguish a vocabulary limitation from a genuine choice between available harmonic alternatives. This distinction helps formulate specific next questions: Would a broader candidate vocabulary suffice? Would revised tonal inference change the options? Would an alternative segmentation better match human harmonic rhythm? Such questions can be tested in controlled subsequent versions without retroactively changing the baseline.

The higher function agreement must be interpreted cautiously. Human chord labels are mapped to functions using the model's detected tonal context; this is not independent validation of a composer's intended functional analysis. Similarly, the maximum-overlap alignment may obscure local chord changes in the 40 mixed-reference segments. These caveats constrain the strength of conclusions about human intention.

Beyond quantitative agreement, the distinction between representational limitations and selection divergence has practical implications for human-centered music systems. For musicians and music educators, an explainable harmonizer can support the exploration of alternative harmonic choices rather than presenting deviations from a reference arrangement as errors. In the 146 segments where the human-reference chord was available but not selected, the system provides opportunities to examine competing harmonic decisions in relation to melodic compatibility, harmonic function, and global progression. Such explanations could support music education, creative experimentation, and human–AI collaboration by encouraging users to compare alternatives and make informed musical decisions. These potential educational benefits, however, remain to be evaluated in future user studies.

### Limitations and future development

The sample contains only 19 processed excerpts, with non-random eligibility constraints and one excluded compound-meter case. The fixed harmonic grid can disagree with human harmonic rhythm; incomplete reference coverage and chord normalization may influence scores. No human listening test, subjective quality assessment, independently validated functional analysis, or causal ablation of scoring terms was conducted. Although a post-hoc comparison with HookTheory key annotations was performed, these annotations were not independently verified against the full musical context. We therefore do not claim that the generated harmony is better or worse, or that differences reveal the original composer's psychology.

A subsequent EMH version will broaden the vocabulary to additional major- and minor-seventh, sixth, secondary-dominant and selected extended chords. Controlled comparisons should separately vary candidate vocabulary, tonal inference, segmentation and decision weights. This would establish whether agreement gains reflect representation, selection or their interaction. The frozen v0.12.1 evaluation provides the baseline against which such changes can be tested.

An additional post-hoc audit compared the melody-derived keys inferred by EMH with the tonal annotations provided by HookTheory. Exact agreement in tonic pitch class and major/minor mode was observed in 8 of the 19 processed excerpts (42.1%). One excerpt shared the same tonic but differed in mode, while three shared the same mode but differed in tonic. The remaining seven differed in both dimensions or involved a non-standard scale annotation. This limited tonal agreement is important when interpreting the reported 47.62% harmonic-function agreement, which is calculated relative to the key inferred by EMH rather than independently annotated human harmonic functions. Tonal ambiguity and differences between melodic and harmonic interpretations may contribute to these discrepancies; however, the present audit does not establish their individual causes.

## CONCLUSION

This study evaluated a frozen explainable melody harmonizer on independently collected HookTheory excerpts, with human chord annotations excluded from the generation process. Across 439 harmonic segments, exact chord agreement reached 20.50%, while chord-root agreement reached 31.66%. Rather than interpreting these differences as simple errors, the decision-level audit distinguished 203 segments in which the reference chord was unavailable from 146 segments in which it was available but not selected.

A supplementary tonal-context audit revealed exact agreement between EMH-inferred and HookTheory-annotated keys in 8 of 19 excerpts (42.1%). This finding highlights tonal inference as an additional factor in interpreting harmonic decisions and cautions against treating the reported 47.62% harmonic-function agreement as an independent measure of human functional intent.

The principal contribution is therefore not the reproduction of a particular human arrangement, but a transparent and reproducible framework for investigating why computational and human harmonizations differ. By distinguishing representational constraints, alternative chord selections, and tonal-context discrepancies, the framework provides a basis for targeted model improvements and for the exploration of harmonic alternatives by musicians and educators. Future work will evaluate these possibilities through controlled model comparisons and human-centered studies.

## ACKNOWLEDGMENT

The authors acknowledge the contributors to the HookTheory dataset and its distribution through the SheetSage project. The Explainable Melody Harmonizer (EMH), experimental scripts, evaluation procedures, and reproducibility materials are publicly available in the Explainable_Melody_Harmonizer directory of the GitHub repository:

https://github.com/JosefPavlicek/quantum-inspired-music-research

The experimental version is preserved under the tag emh-v0.12.1-hooktheory. Dataset access and reuse remain subject to the original licensing and distribution terms.

Generative AI tools were used to assist with English-language editing, grammatical correction, stylistic refinement, and the preparation of illustrative graphical materials. The authors retain full responsibility for the scientific content, methodological validity, verification of results, interpretations, and final conclusions presented in this work.